\documentclass[conference]{IEEEtran}
\IEEEoverridecommandlockouts
\usepackage{cite}
\usepackage{amsmath,amssymb,amsfonts}
\usepackage{algorithmic}
\usepackage{graphicx}
\usepackage{textcomp}
\usepackage{xcolor}
\usepackage{listings}
\def\BibTeX{{\rm B\kern-.05em{\sc i\kern-.025em b}\kern-.08em
    T\kern-.1667em\lower.7ex\hbox{E}\kern-.125emX}}

\begin{document}

\title{Automated Driver Testing for Small Footprint Embedded Systems\\
\thanks{This work was partially funded by Fundação de Amparo à Pesquisa do Estado do Rio de Janeiro (Faperj) and Conselho Nacional de Desenvolvimento Científico e Tecnológico (CNPq).}
}

\author{\IEEEauthorblockN{Sara C. M. Souza}
\IEEEauthorblockA{\textit{Innovation Hub} \\
\textit{Instituto Federal Fluminense, Brazil}\\
ORCID: 0000-0002-0905-8650}
\and
\IEEEauthorblockN{Rogerio Atem de Carvalho}
\IEEEauthorblockA{\textit{Innovation Hub} \\
\textit{Instituto Federal Fluminense, Brazil}\\
ORCID: 0000-0003-4429-2482}
}

\maketitle

\section{Introduction}

Embedded systems are present on many applications such as in automotive, aerospace, industrial, and medical applications. These systems drive a billionaire market and they are responsible for more than 90\% of the processors produced in the world \cite{banerjee2016testing}.

Embedded System are a combination of hardware and software designed for a specific purpose, with multiple software layers. Drivers are part of a low level software layer intended to interact directly to hardware. For example, the internal peripherals like Timer Modules, SPI Modules and so on. So, there is a strong interaction between the drivers and internal or external peripherals and this interaction often makes the embedded software development, functionality evaluation, system performance measurement, and automated testing a hard task\cite{grenning2011test}. This happens because testing demands control over the inputs and outputs, in this case the hardware, this task is especially complex for embedded systems given that the physical environment may not be deterministic and difficult to be recreated during the tests \cite{banerjee2016testing}. For example, in order to test a driver which is responsible to produce a PWM output signal it may be necessary a device in the other side that is able to capture this signal and verify the frequency of this signal, or to test a driver that is responsible to get a GPS data it may be necessary to provide a device that answer a predefined GPS data requests and uses the same physical communication protocol. 

Although in many cases a developer may use real external peripherals to evaluate his/her driver, in other cases the hardware may not be available during the software development because it is also in development or because of its high price \cite{grenning2011test}. Moreover, the hardware presence may not be wished in some situations because of the risks it may represent \cite{ellis2012control}. To understand how the testing task is relevant and costly, a worldwide research \cite{vincenzi2018automatizaccao} indicated that 40\% of the assigned resources on a project are applied in software testing activities and can reach up to 80\% on critical systems.

In order to address the problems previously described, and taking into account that a great deal of embedded systems, especially IoT devices, are based on microcontrollers or SoCs, this paper introduces a low cost solution to test drivers on this type of computing element. The solution is built upon a method that employs three components: a Device Under Test (DUT), a Double device, and a personal computer. The computer runs the test code, i.e., the code that will compare the expected result with the obtained result, while the DUT runs the test's target code, i.e., the driver that the developer wishes to test, and the Double plays the role of the "real" external peripherals that communicates with the DUT, is this device that the driver is supposed to control or communicate with.

The history of this solution begins with the control software for one of the payloads of the Brazilian nanosatellite 14-BISat. The development of this instrument ran in parallel with the development of its control software, which was carried out by the authors, who used basic Test Driven Development techniques and a test harness called THC \cite{carvalho2014, carvalho2015}. These initial experiments generated the Valves 1.0 toolset, which in turn was used to certify the embedded software in 14-BISat, as described in \cite{carvalho2016a, carvalho2016b, carvalho2017}. From this and other accumulated experiences and developments, different requirements were collected in order to improve the process of testing embedded systems, among which, the creation of hardware and software Doubles, which gave rise to the solution here presented.

In order to briefly present a solution matured over the years to test the drivers from small footprint embedded devices, this article is divided into the following sections:
\begin{itemize}
    \item[-] The proposed solution section, which presents the system architecture and the method in detail;
    \item[-] The test cases section, that describes driver tests performed using up to five different communication protocols;
    \item[-] Conclusions, which list the main achievements of the solution, as well as its shortcomings and current directions.
\end{itemize}

\section{Proposed Solution}
Here is introduced a testing solution for drivers on microcontrollers that, besides the Device Under Test (DUT), has a a Double device, and a computer, as well as software associated to each of these components. Their structure and interactions are described as follows.

\subsection{System Architecture}

\begin{figure}[htbp]
\centerline{\includegraphics[scale=0.35]{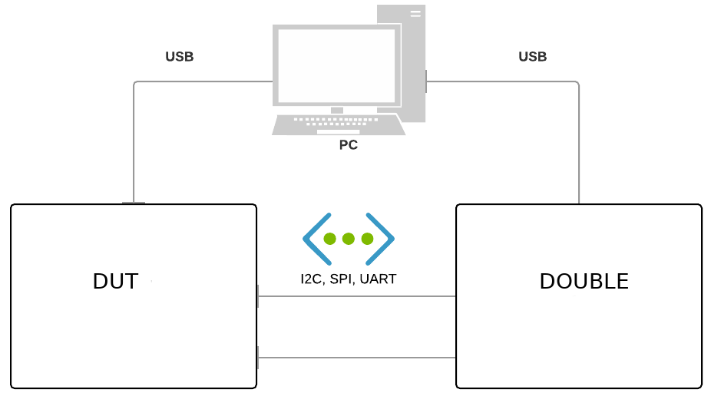}}
\caption{Hardware elements.}
\label{devices}
\end{figure}

\begin{figure}[htbp]
\centerline{\includegraphics[scale=0.35]{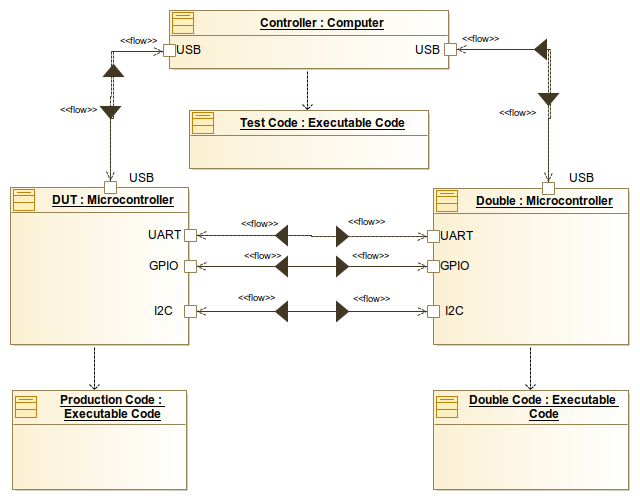}}
\caption{System Architecture.}
\label{architecture}
\end{figure}

The hardware elements are represented in Fig.~\ref{devices}. As can be seen in Fig.~\ref{architecture}, System Architecture, each hardware element hosts a software element. The hardware devices are described as follows:

\begin{enumerate}
    \item Computer: This device hosts the test code and it is responsible for orchestrating the test by sending commands to the DUT and to the Double, and for receiving the results of these commands from both;
    \item Device Under Test: This device is a microcontroller or SoC whose driver is being tested or evaluated. This device hosts the test's target code, which is called the production code;
    \item Double Device: A Double device is a device that plays the role of an external peripheral, in the same way that a double plays the role of a real actor on a movie. The Double provides inputs to the device under test and receives outputs from it, making the doubled devices transparent from the DUT’s viewpoint. It supports multiple protocols in order to allow the testing of different drivers. The Double device hosts the so called Double Code.
\end{enumerate}

The software elements are:
\begin{enumerate}
    \item Test Code: it is a Python code running in the computer. It orchestrates the test by communicating to each device through the injection of commands, and gathering results from a serial connection. Once the outcomes are gathered, this code compares the gathered results against the expected results and gives feedback to the developer. A test code may have one or more test cases. Each test case aims at testing a small and isolated feature, such as a read/write operation. According to the Test Driven Development (TDD) philosophy, this code should be interactively written before the production code;
    \item Double Code: This code is embedded into the Double device and simulates the behavior of the real devices. A Double code usually declares a class that represents an external peripheral, this way the developer may instantiate one or more testing objects, such as leds from the Led Class, for example. The solution here presented already supplies some functional Doubles, however, the development community and the peripheral manufacturers can produce their own Double Code, supplying extra doubles as necessary;
    \item Production Code: The code hosted by the DUT that is the target of the tests. It implements the drivers. This code may be any embedded code.
\end{enumerate}

\subsection{Materials and Methods}

Currently, the Materials used for testing are:
\begin{itemize}
    \item Two DevKit v1 Evaluation Board (one being the DUT and the other being the Double). This board host the ESP32, a System On a Chip (SoC) that was chosen because of its many protocols, the BLE and WiFi feature and because it is very accessible.
    \item One laptop running Ubuntu 18.04 Linux Distro. 
\end{itemize}

Regarding the Method, it is centered in what is called a complete test execution, which is divided into two phases:
\begin{itemize}
    \item Setup phase: This phase is responsible for establishing a connection, cleaning up the DUT's and Double's file systems, and uploading the proper codes to each device. 
    \item Execution phase: It is responsible for running all test cases, when each test case should evaluate a single functionality.
\end{itemize}

\begin{figure}[htbp]
\centerline{\includegraphics[scale=0.35]{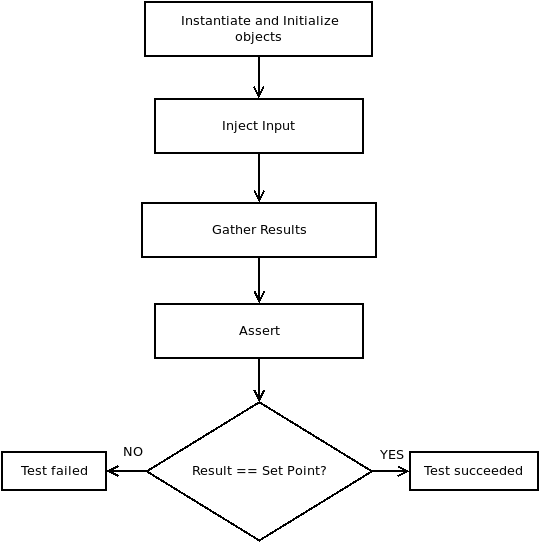}}
\caption{Execution flow of a unique test case.}
\label{testcaseflow}
\end{figure}

Fig.~\ref{testcaseflow} represents the execution flow of a generic test case. These steps are described as follows: 

\begin{enumerate}
    \item 
Initialization: the Test Code initializes the production code in the DUT and the Double Code in the Double device. This is achieved through the instantiation and initialization of objects on each device using the serial communication previously established. This step is responsible, for example, for instantiating a simulated object on the Double device and for indicating which pins it should use. 
\item
Input Injection: the Test Code places the production code at a state in which the developer wants to test, also placing the Double Code at a suitable state corresponding to that test case. A suitable state in this case means a favorable state for the test case considered, e.g., making the Double sensitive to external stimulus or making it produce an output. This state is based upon the real behavior of the given peripheral.
\item
Results Gathering: the Test Code requests the results from the DUT or the Double Device. 
\item
Asserts: the Test Code compares the results with the set-point. The outcome of this assertion is printed out, indicating the result of the test case itself. This phase uses a Python Module called \verb|should dsl| \cite{carvalho2010} that allows the use of specific \textit{matchers}. \textit{Matchers} are more than just basic asserts, they allow more sophisticated comparisons, such as specifying a range of tolerance, asserting if, for example, a test is successful if a timer expiration value is between 1999ms and 2001ms.   
\end{enumerate}

For standardization purposes, it was established that Test Codes' names should initialize with \verb|"test_"|, production codes with \verb|"dut_"| and Double Codes with \verb|"Double_"|, followed by an expressive name indicating the main goal of the code.

\section{Cases}

This section presents and discuss the currently implemented cases. Each subsection aims at exemplifying how to apply the method to test a specific driver that uses a specific internal peripheral and protocol. For example, the first subsection presents the Blink Led case, describing how to test the driver responsible for blinking a led, using the microcontroller's General Purpose Input/Output (GPIO) as the internal peripheral. Each subsection presents the following topics:
\begin{enumerate}
    \item A brief description of the test cases. 
    \item A representative test case snippet.
    \item The test result.
    \item The electrical connection between the DUT and the Double. 
\end{enumerate}

\subsection{The Blink Led case}

\begin{enumerate}
    \item Brief Description: when a developer starts programming in a language, it's common to start with the classic "Hello World" printing code. In the microcontroller's environment, the blink led code is often used as the first exercise in embedded systems studies, as well as to check the basic functioning of a new evaluation board. Therefore, the classic Blink Led was used to test GPIO and also as an introduction to the proposed solution's method.
    \item Test Cases: The \verb|"test_blink_led.py"|, which is the test code, has the following test cases: The \verb|"test_blink_blocking"| and the \verb|"test_blink_isr"|. The former is intended to test a blink code operating in the blocking mode, i.e., consuming CPU's cycles, and the latter is intended to test a blink code that uses a timer interrupt service routine to blink the led. 
    The \verb|"test_blink_blocking"| case is exposed in Code~\ref{test_blink_blocking}. The beginning of this code is composed by constants that describe, respectively, in which pin the led will be connected to the DUT, in which pin the Double will pretend to be the led, i.e., the pin the Double will receive the stimulus, how long will the blinking period last and how many times it will blink.
The First part is the object initialization, with the test initially creating the Led object in the Double, and then sending the command in the string format via the serial connection previously established. The command indicates to the Double which pin it will listen to, and after how many times it can stop and return to the test the average blink time. Then the test creates the Blinker object in the DUT, specifying a pin to blink, a blinking period, and the number of blinks. It's important to emphasize that the test code had already sent the class file to these devices before the instantiating process. Also, the Blinker object contains the driver itself which is the test's target.
The second part is the input injection, which in this case consisted of calling a Led's method to start the signal acquisition and of calling a Blinker's method to finally blink the led. In the third part, the test sleeps, in order to wait the led to blink, and after this interval, the test asks the Double for the blink average time, which is, in that case, the result.
The fourth part is the assertion. Here, it's interesting to observe that it was used the \verb|"close_to"| matcher, which gives the developer the possibility of giving the test a tolerance. In this example, the maximum absolute error was of 1 millisecond. It means that the test passes if the time period is 1 millisecond more or less. 
The last part was simply the decommissioning, i.e., memory cleaning. The \verb|"test_blink_isr"| is pretty similar to this case, with the difference in the called method on DUT, in which a timer interrupt service routine is configured. 

\end{enumerate}

\begin{lstlisting}[
language=Python, 
caption={Test case: test\_blink\_ blocking()},
label={test_blink_blocking},
frame=single]
led_pin_dut = "2"
led_pin_double = "21"
period_blink = "1000"
blinking_times = "2"
...
# 1 - Objects Creation
self.double_serial.repl("red_led =
Led("+led_pin_double+",
"+blinking_times+")",0.1)
self.dut_serial.repl("blinker =
Blinker("+led_pin_dut+","+period_blink+",
"+blinking_times+")",0.1)

# 2 - Input Injection
self.double_serial.repl(
"red_led.start_acquisition()",0.1)
self.dut_serial.repl(
"blinker.blink_blocking()",0.1)

# 3 - Results gathering
sleep(...) #To give the test enough time
obtained_period = self.double_serial.repl(
"red_led.get_average_period()",0.2)[2]
...

# 4 - Assertion
obtained_period |should| close_to 
(period_blink, delta = 1)

# 5 - decommissioning
self.dut_serial.repl("del blinker",0.1)
self.double_serial.repl("del red_led",0.1)
\end{lstlisting}

\begin{figure}[htbp]
\centerline{\includegraphics[scale=0.27]{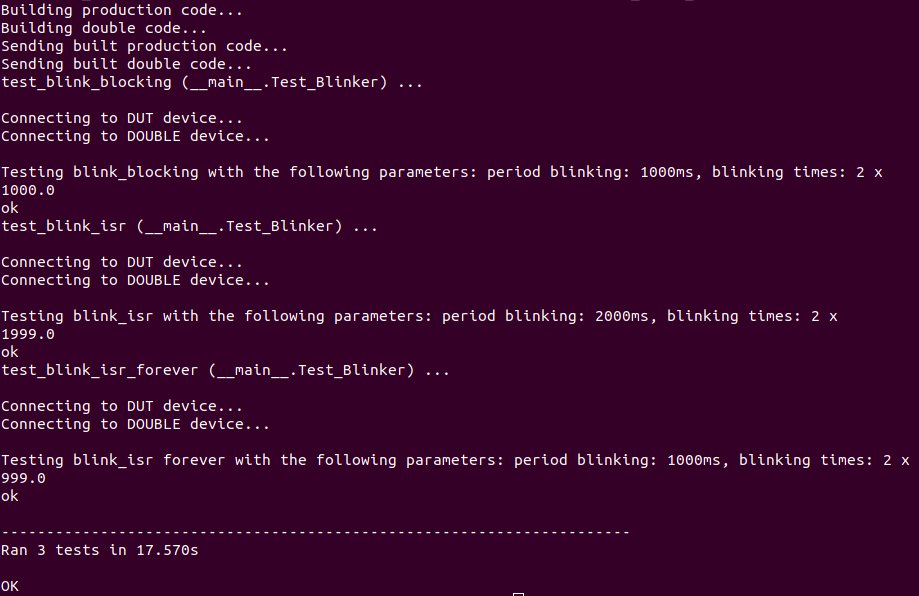}}
\caption{The blink test result.}
\label{result_gpio}
\end{figure}

\begin{enumerate}
    \setcounter{enumi}{2}
    \item Results: Fig.~\ref{result_gpio} presents the result of the Blink Led test, gives detailed feedback about each test case and provides the input test parameters and the output.
\end{enumerate}

\begin{figure}[htbp]
\centerline{\includegraphics[scale=0.45]{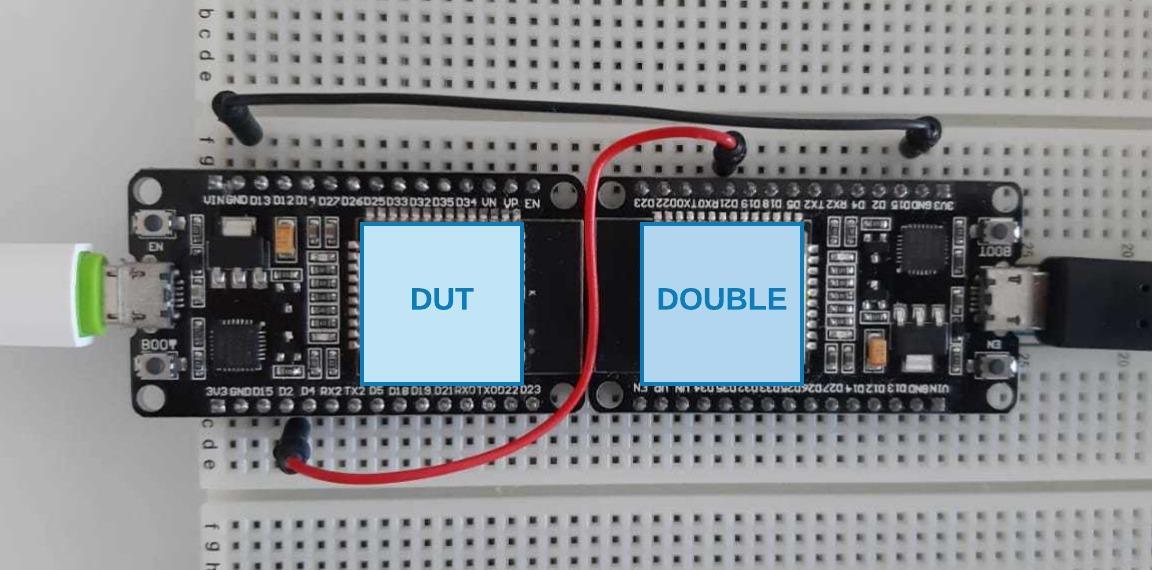}}
\caption{The blink test electrical connection.}
\label{ligacao_gpio}
\end{figure}

\begin{enumerate}
    \setcounter{enumi}{3}
    \item The electrical connection: Fig.~\ref{ligacao_gpio} presents the electrical connection for testing the blink led. There are two wires, one for the led's fake connection and one for the ground.
\end{enumerate}

\subsection{The RTC case}

\begin{enumerate}
    \item Brief Description: An external Real-Time Clock (RTC) is often used to keep the Date and Time information even when the device is turned off. The RTC of this example uses an I2C interface to communicate with the microcontroller. The basic expected functions of an RTC's driver are setting and getting the Date and Time from the RTC. Therefore, this test aimed at separately testing each of these functionalities. The isolated test for these functionalities deserves special attention because this approach allows the \verb|get| method to be tested even when the \verb|set| method is not implemented. To do so, the Double plays the role of an RTC model DS3231.  
    \item Test Cases: This test set contains four test cases. The \verb|test_set_date_time_static| is intended to test the set method, the static suffix is used because in this case the Double keeps the Date and Time instead of start counting from that Date and Time onward. However, the \verb|test_set_date_time_dynamic| is also intended to test the set method, in this case, the Double operates in the dynamic mode, i.e., it creates fake registers where the Date and Time information is stored and it uses its internal timer to update these registers. This mode brings a more realistic behavior and allows a dynamic testing experience. This case is exposed in  Code~\ref{test_set_date_time_dynamic}. The \verb|test_get_date_time| is simply intended to test the get method. And the case \verb|test_set_get_date_time| is intended to test a full operation: a set call followed by a get call. This last case was specially designed to accomplish an autotest, i.e., the DUT injects the input through the get method and gathers the result through the set method, and verifies the functioning of its own external peripheral before operating.
\end{enumerate}

\begin{lstlisting}[
language=Python, 
caption={Test case: test\_set\_date\_time\_dynamic()},
label={test_set_date_time_dynamic},
frame=single]
 # datatime format: [Year, month, day,
# weekday, hour, minute, second]
expected_datetime =[2019,10,23,4,10,51,32]
my_delay = 2

# 1 - Objects Creation
self.double_serial.repl(
"ds3231 = DOUBLE_DS3231(21,22)",0.2)
self.dut_serial.repl(
"rtc_conf = ConfTime(21,22)",0.2)

# 2 - Input Injection
self.dut_serial.repl(
"rtc_conf.set_date_time(
"+str(expected_datetime)+")", 0.4)
self.double_serial.repl(
"ds3231.use_internal_rtc()", 0.2)
sleep(my_delay)
expected_datetime[6] += my_delay 

# 3 - Results gathering
gotten_datetime = self.double_serial.
repl("ds3231.DateTime()",0.2)[2]

# 4 - Assertion
...
gotten_datetime |should| equal_to
(str(expected_datetime))
\end{lstlisting}

\begin{figure}[htbp]
\centerline{\includegraphics[scale=0.40]{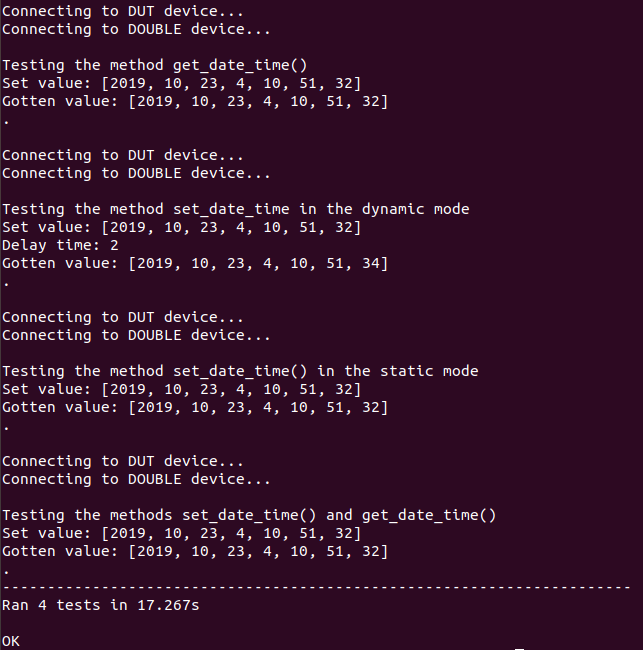}}
\caption{The conf time result.}
\label{result_i2c}
\end{figure}

\begin{enumerate}
    \setcounter{enumi}{2}
    \item Results: Fig.~\ref{result_i2c} presents the result of the Conf Time test, giving detailed feedback about each test case, providing the input test parameters and the output.
\end{enumerate}

\begin{figure}[htbp]
\centerline{\includegraphics[scale=0.20]{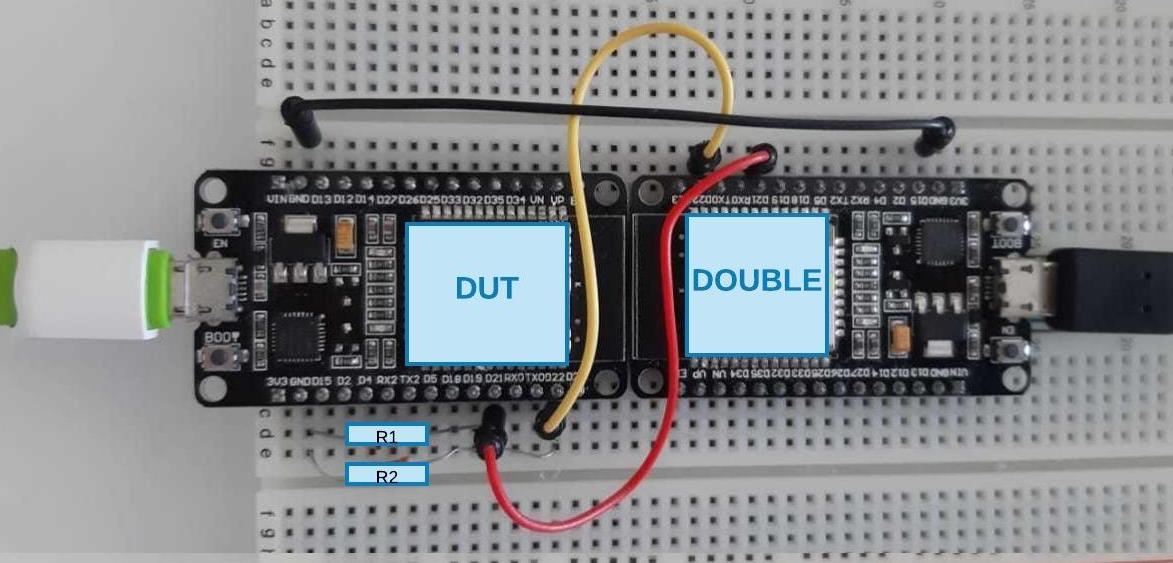}}
\caption{The conf time test electrical connection.}
\label{ligacao_i2c}
\end{figure}

\begin{enumerate}
    \setcounter{enumi}{3}
    \item The electrical connection: Fig.~\ref{ligacao_i2c} presents the electrical connection for the RTC's testing, having three wires, two for the I2C connection and one for the ground. There are also two pull-up resistors. 
\end{enumerate}

\subsection{The GPS case}

\begin{enumerate}
    \item Brief Description: An external Global Positioning System (GPS) is often used to get geographic coordinates, date, time, and other information. The GPS module of this example uses a UART interface to communicate with the microcontroller. The Double was based on the model NEO-6M, a very popular module in the microcontrollers' arena. A microcontroller configures the GPS module and receives the information from it through National Marine Electronics Association (NMEA) sentences. NMEA sentences hold specific codes and a format to identify the configuration or the information \cite{nmea_codes}. Although the major part of GPS libraries are composed of parser and handler methods used to interpret these NMEA sentences, the test here presented focuses only on sending commands and the reception of related information, given that this is the part of the library that interacts with the hardware. 
    \item Test Cases: This test contains three test cases. The \verb|test_send_command_configuration| is intended to test the \verb|send_command| method, which is responsible for configuring the module by sending a NMEA sentence that indicates the kind of information the microcontroller wants to receive. 
    The \verb|test_send_command_update_rate| also sends a command, but this time the command indicates to the GPS module the update rate in which it should send the information.  When the Double receives such a command, it enables a timer to periodically send a preconfigured NMEA sentence, e.g., a static sentence that contains a latitude and longitude value. This possibility is especially useful when a developer is creating the driver in an environment in which satellite signals are weak or there are other connection problems - in special laboratories where signal interference is quite common. This case is exposed in the Code~\ref{test_send_command_update_rate}. The \verb|test_get_latitude| is intended to test the multiple library layers, from the driver up to the parser and handler. For this purpose the \verb|get_latitude| method is tested and it includes the UART reading and the interpretation of the received sentence.
\end{enumerate}

\begin{lstlisting}[
language=Python, 
caption={Test case: test\_send\_command\_update\_rate().},
label={test_send_command_update_rate},
frame=single]
command = 'PMTK220,1000'
expected_answer = '$GPRMC,
141623.523,A,2143.963,S,04111.493,
W,,,301019,000.0,W*7B\n'

# 1 - Objects Creation
self.double_serial.repl(
"gps = DOUBLE_GPS(22,16,2)",0.2)
self.dut_serial.repl(
"tracker = Positioning(17,16,2)",0.2)

# 2 - Input Injection
self.dut_serial.repl(
"tracker.send_command('"+command+"')",0.2)
self.double_serial.repl(
"gps.received_command()",0.2)
sleep(1)

# 3 - Results gathering
gotten_answer = self.dut_serial.repl(
"tracker.received_command()",0.2)[2]

# 4 - Assertion
...
gotten_answer |should| equal_to
(expected_answer)
\end{lstlisting}

\begin{figure}[htbp]
\centerline{\includegraphics[scale=0.35]{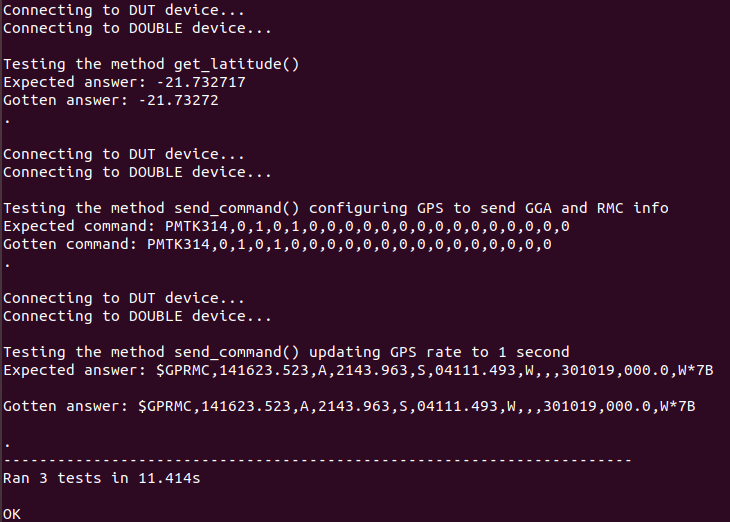}}
\caption{The positioning GPS test result.}
\label{result_uart}
\end{figure}

\begin{enumerate}
    \setcounter{enumi}{2}
    \item Results: Fig.~\ref{result_uart} presents the result of the Positioning GPS test, giving detailed feedback about each test case, providing the input test parameters and the corresponding output.
\end{enumerate}

\begin{figure}[htbp]
\centerline{\includegraphics[scale=0.20]{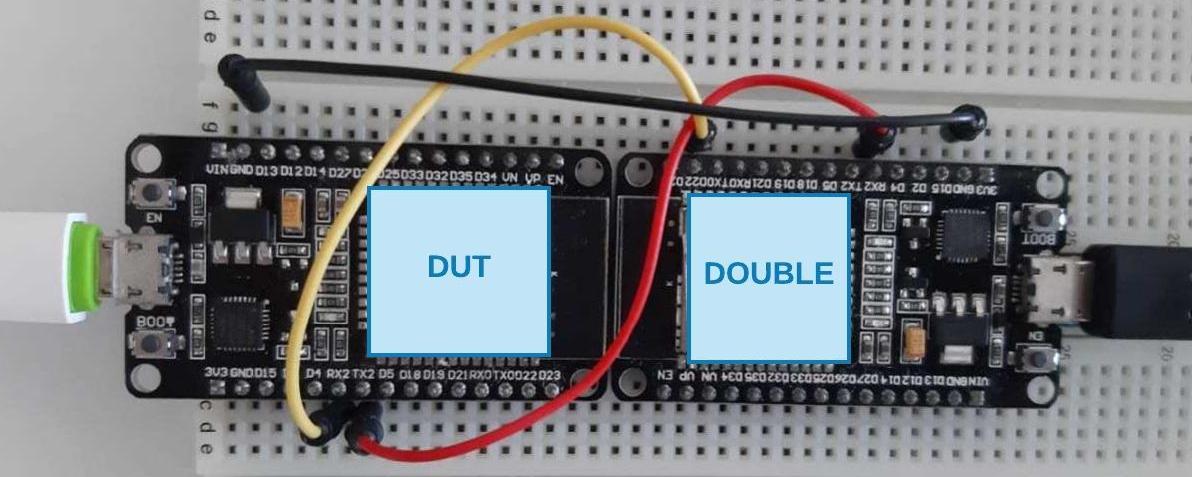}}
\caption{The Positioning GPS test electrical connection.}
\label{ligacao_uart}
\end{figure}

\begin{enumerate}
    \setcounter{enumi}{3}
    \item The electrical connection: Fig.~\ref{ligacao_uart} presents the electrical connection for the UART testing, having three wires, two for the TX and RX lines, and one for the ground.
\end{enumerate}

\subsection{The SPI case}

\begin{enumerate}
    \item Brief Description: This example shows how to test the operation of sending and receiving data on the SPI bus. The DOUBLE is playing the role of a generic SPI slave as any peripheral that uses SPI to talk to a microcontroller. After stress testing, meaning repeatedly executing this driver's tests, the test revealed an instability of the SPI protocol when operating in the slave mode, i.e., the Double's SPI. Thus, although the test target was the DUT, which works in the master mode, the test results exposed an instability of SPI working in slave mode. Instability here means that sometimes the test passed and other times the test failed, unpredictably.
    \item Test Cases: The test cases basically consisted on sending and receiving data to the DUT. To test the send operation, the data received by the Double was compared with the data sent by the master (DUT). To test the receive operation, the data sent by the Double was compared to the data received by the DUT. To exemplify with one test case, Code~\ref{test_reading_registers_without_indicating_address} exposes a case that was intended to perform a read operation without indicating a register address, i.e., only retrieving the available data in the RX buffer, exactly like the read operations in ADCs.

\end{enumerate}

\begin{lstlisting}[
language=Python, 
caption={Test case: test\_reading\_registers\_without\_indicating\_address().},
label={test_reading_registers_without_indicating_address},
frame=single]
expected_written = [1,2,3]
gotten_values = []

# 1 - Objects Creation
self.double_serial.repl
("slave = Double_slave(13,12,14,15)",0.2)
self.dut_serial.repl
("master = Dut_master(13,12,14,15)",0.2)

# 2 - Input Injection 
self.double_serial.repl
("slave.enable_transaction([0,0,0])",0.2)
self.dut_serial.repl
("master.write_registers_wo_address
("+str(expected_written)+")",0.2)

# 3 - Results gathering
gotten_values = self.double_serial.repl
("slave.get_received_buffer()", 0.2)[2]
gotten_values = gotten_values.decode()

# 4 - Assertion
gotten_values |should| equal_to
 (str(expected_written))
\end{lstlisting}

\begin{figure}[htbp]
\centerline{\includegraphics[scale=0.35]{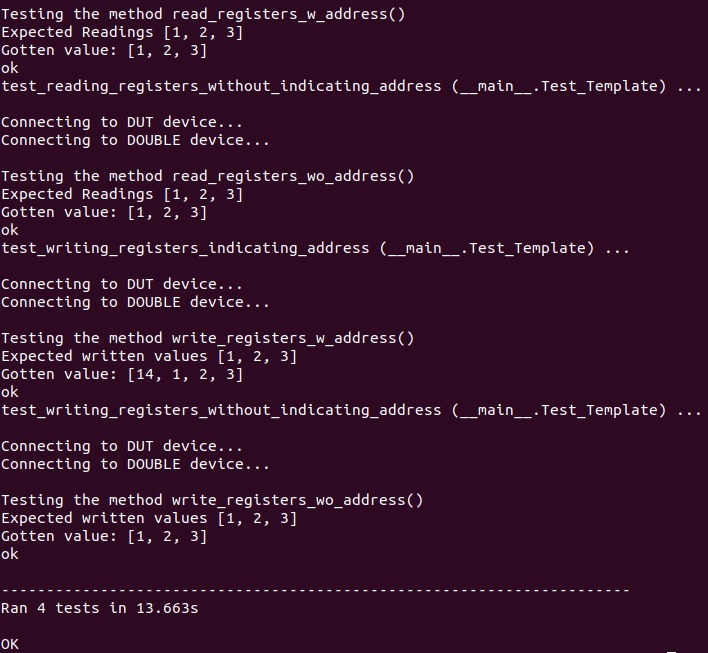}}
\caption{A successfull SPI test result.}
\label{result_spi}
\end{figure}

\begin{figure}[htbp]
\centerline{\includegraphics[scale=0.30]{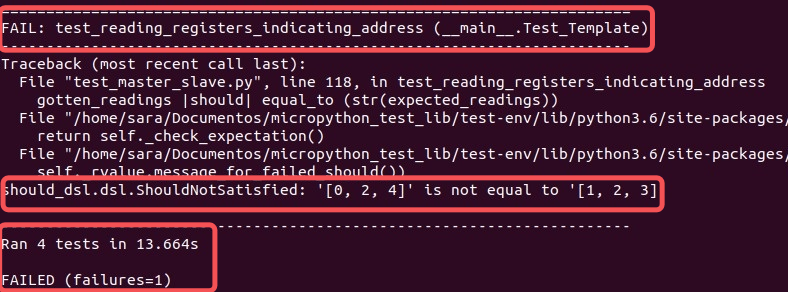}}
\caption{A failed SPI test result.}
\label{result_spi2}
\end{figure}

\begin{enumerate}
    \setcounter{enumi}{2}
    \item Results: Fig.~\ref{result_spi} presents the result of a successful SPI test while Fig.~\ref{result_spi2} presents a failed execution.
\end{enumerate}

\begin{figure}[htbp]
\centerline{\includegraphics[scale=0.40]{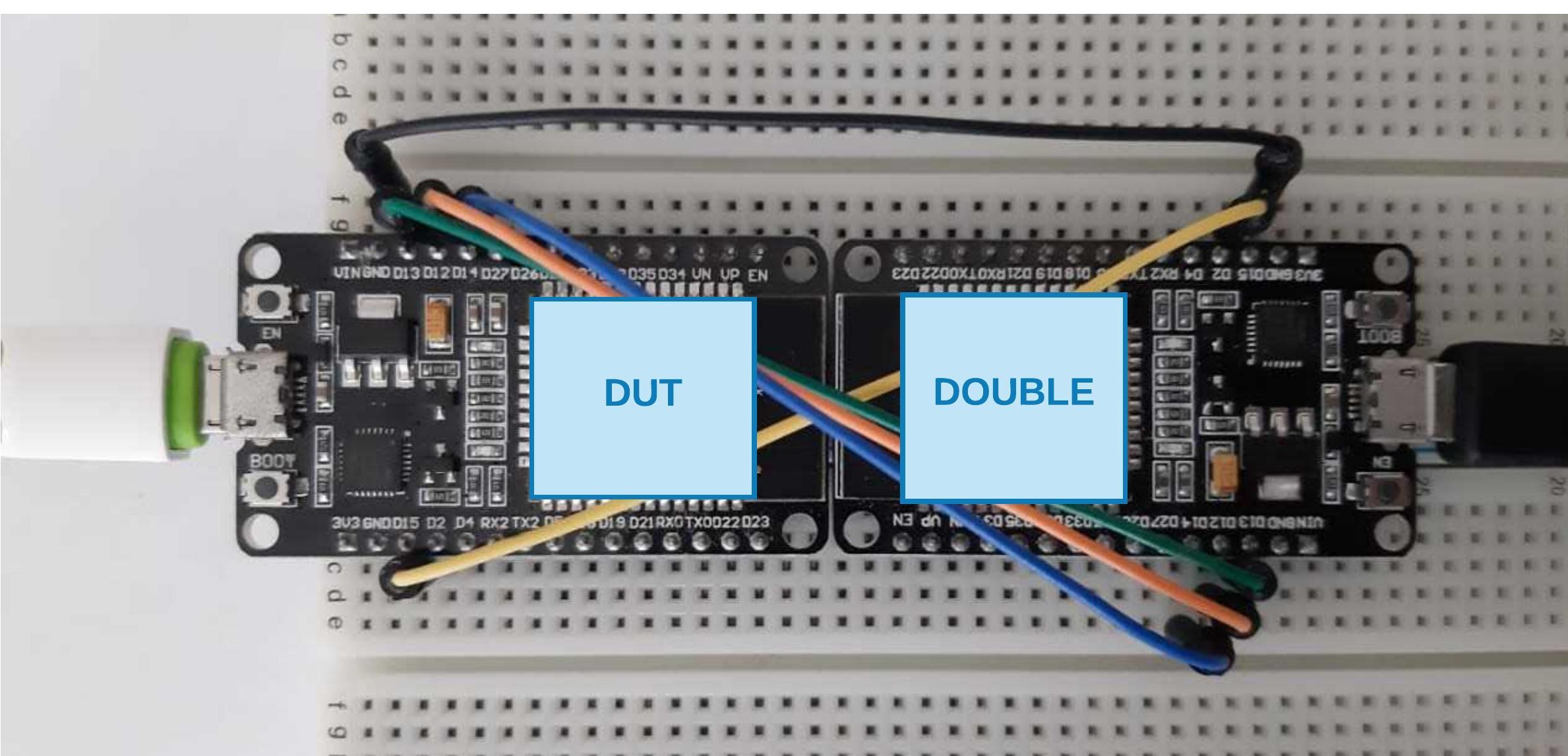}}
\caption{The SPI test electrical connection.}
\label{ligacao_spi}
\end{figure}

\begin{enumerate}
    \setcounter{enumi}{3}
    \item The electrical connection: Fig.~\ref{ligacao_spi} presents the electrical connection for the SPI testing, having five wires: four for the SPI communication (SCLK, MISO, MOSI and CS) buses, and one for the ground.
\end{enumerate}

\subsection{The BLE case}

\begin{enumerate}
    \item Brief Description: The Bluetooth Low Energy is a wireless communication protocol introduced by the Bluetooth 4.0 specification. It's often used in MCU, sensors, smart watches and IoT devices to communicate to other more robust devices such as smartphones and laptops. The BLE specification uses the Generic Access Profile (GAP) and Generic Attribute Profile (GATT) standards. Together, these standards specifies roles played by the devices, how they discover each other, how they establish a connection, and how they change information.  According to the GAP specification, before establishing a connection, a Peripheral device starts advertising its offered services, address, and other informations. Thus, a Central device will be able to scan these information and initialize a connection. Small devices, like IoT devices, often operates in the Peripheral mode, while robust devices, like smartphones, often operates in the Central mode \cite{ble}.   
    In this test, the DUT is operating in the Peripheral mode, playing the role of a generic Temperature Sensor and the goal was to test the code that allows a connection and reads and notifies the temperature over Bluetooth. To accomplish it, the Double operated in the Central mode and played the role of a smartphone.
    \item Test Cases: This test set contains three test cases. The \verb|test_connection| intended to test the connectivity between the sensor and the smartphone.
    The \verb|test_read| and \verb|test_notify| intended to test a read operation initiated by the smartphone and the notification promoted by the sensor. To exemplify with one test case, Code~\ref{test_connection} exposes the connection test case.
    
\end{enumerate}

\begin{lstlisting}[
language=Python, 
caption={Test case: test\_connection().},
label={test_connection},
frame=single]
advertising_name = "my_sensor"

# 1 - Objects Creation and Input Injection

self.dut_serial.repl("sensor = BLE_Sensor
(\""+advertising_name+"\")", 0.2)
self.double_serial.repl("cellphone =
DOUBLE_cellphone()", 0.2)
time.sleep(5)

# 3 - Result Gathering

connection = self.double_serial.repl
("cellphone.check_connectivity()", 0.2)[2]
connection = connection.decode()
connection |should| contain (str(True))
\end{lstlisting}

\begin{figure}[htbp]
\centerline{\includegraphics[scale=0.35]{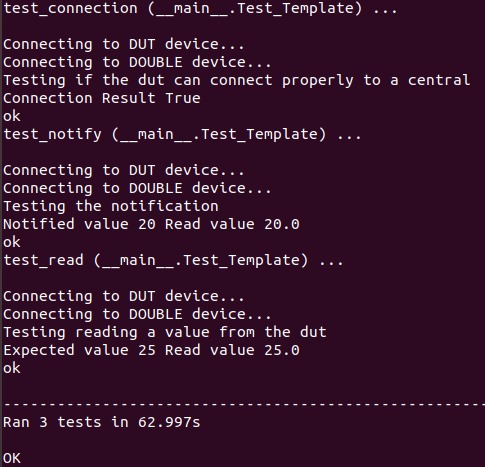}}
\caption{A successfull BLE test result.}
\label{result_ble}
\end{figure}

\begin{figure}[htbp]
\centerline{\includegraphics[scale=0.30]{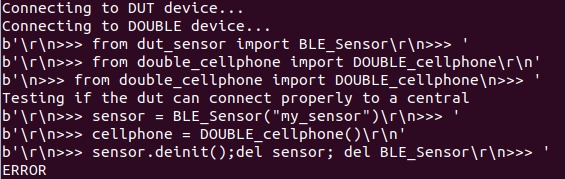}}
\caption{A failed BLE test result.}
\label{result_ble2}
\end{figure}

\begin{enumerate}
    \setcounter{enumi}{2}
    \item Results: Fig.~\ref{result_ble} presents the result of a successful BLE test while Fig.~\ref{result_ble2} presents a failed execution in debug mode, where it's possible to see the commands being executed in both devices. The test result was unstable given that in some executions the time for BLE initialization was so long that the test was executed before the Double had the chance to finish initializing and establishing a connection. It's possible to realize in Fig.~\ref{result_ble2} that the cellphone was not de-initialized because the Double terminal was blocked still waiting to initialize the BLE. 
    The interesting point is that after we had this result, we searched for the problem and we could find an issue in GitHub regarding it \cite{blebug1}. So, once more, the framework showed up effective to reveal driver's implementation problems.
\end{enumerate}

\section{Conclusions}

This paper introduced a solution that deliveries a practical and efficient way of testing drivers in Small Footprint Embedded Systems. The solution is composed by multiple artifacts, including libraries, templates, and test cases, all functional in the given protocols. Among the five protocols here presented, Bluetooth Low Energy (BLE) and Serial Peripheral Interface (SPI) have shown instability, due to the recent libraries' development for the target platform. However, an interesting result obtained during the development of the solution was that the tests created for these protocols were capable of exposing these instabilities, which were then reported to the corresponding development communities. All the artifacts, along with clear instructions are published in GitHub at \cite{myrepo}, as a invitation to collaborators. 

The target platform and language of choice, ESP32 and Micropython, were chosen due to the protocol diversity of this SoC and the possibility of accessing an interactive terminal that allows running the tests and reading their results, respectively. It is important to note that the Double can be used with other platforms, given that it implemented in a transparent way - exactly to act as a double. 

Also, it must be observed that the method is noninvasive, which means that there is no need for changing the production code or loading code into the target hardware in order to going under testing. There's no breakpoints and no necessity for setting a special debug mode for running the tests. However, in practice, the Test Code's execution may add some delays when sending commands and receiving results through the serial port, reinforcing the need for good test design.

Therefore, our contribution is threefold: studying the extension of the concept of Doubles from software to hardware, supplying a functional hardware and software solution for testing drivers in Small Footprint Embedded Systems, and pinpointing errors in two of driver implementations for the developer community. Current work is divided into exploring other protocols and providing test-case automated code generation. 

\bibliographystyle{IEEEtran}
\bibliography{myreferences}

\end{document}